\documentclass{article}
\usepackage{amsmath}
\usepackage[psamsfonts]{amssymb}
\usepackage{cmmib57}

\def\QED{\hskip0.1em\hfill\null\ \null\nobreak\hfill\kern3pt\vbox{\hrule\hbox
   {\vrule\kern1pt\vbox{\kern1.7pt\hbox{$\scriptscriptstyle{QED}$}
    \kern0.2pt}\kern1pt\vrule}\hrule}}

\def\END{\hskip0.1em\hfill\null\ \null\nobreak\hfill\kern3pt\vbox{\hrule\hbox
   {\vrule\kern1pt\vbox{\kern1.7pt\hbox{$\,\,\,\vspace{5pt}$}
    \kern0.2pt}\kern1pt\vrule}\hrule}}
\newtheorem{theorem}{Theorem}
\newtheorem{lemma}{Lemma}
\newtheorem{corollary}{Corollary}
\newtheorem{proposition}{Proposition}
\newtheorem{remark}{Remark}
\newtheorem{definition}{Definition}
\newtheorem{example}{Example}
\newcommand{\bCd}{\bEq\begin{CD}}
\newcommand{\eCd}{\end{CD}\eEq}
\newcommand{\bcd}{\beq\begin{CD}}
\newcommand{\ecd}{\end{CD}\eeq}
\newcommand{\ben}{\begin{enumerate}}
\newcommand{\een}{\end{enumerate}}
\newcommand{\bEq}{\begin{eqnarray}}
\newcommand{\eEq}{\end{eqnarray}}
\newcommand{\beq}{\begin{eqnarray*}}
\newcommand{\eeq}{\end{eqnarray*}}
\newcommand{\bDf}{\begin{definition}\em}
\newcommand{\eDf}{\end{definition}}
\newcommand{\bLm}{\begin{lemma}}
\newcommand{\eLm}{\end{lemma}}
\newcommand{\bPr}{\begin{proposition}}
\newcommand{\ePr}{\end{proposition}}
\newcommand{\bTh}{\begin{theorem}}
\newcommand{\eTh}{\end{theorem}}
\newcommand{\bCr}{\begin{corollary}}
\newcommand{\eCr}{\end{corollary}}
\newcommand{\bRm}{\begin{remark}\em}
\newcommand{\eRm}{\end{remark}}
\newcommand{\bEx}{\begin{example}\em}
\newcommand{\bex}{\begin{example*}\em}
\newcommand{\eex}{\end{example*}}
\newcommand{\eEx}{\end{example}}
\newcommand{\C}{\mathbb{C}}

\newcommand{\ie}{{\em i.e$.$} }
\newcommand{\eg}{{\em e.g$.$} }
\newcommand{\R}{I\!\!R}

\newcommand{\mto}{\mapsto}

\newcommand{\der}{\partial}

\DeclareMathOperator{\Aut}{Aut}
\DeclareMathOperator{\Diff}{Diff}

\DeclareMathOperator{\ad}{ad}

\newcommand{\ucar}[1]{\underset{#1}{\times}}

\newcommand{\cA}{\mathcal{A}}

\newcommand{\cC}{\mathcal{C}}
\newcommand{\cD}{\mathcal{D}}
\newcommand{\cE}{\mathcal{E}}

\newcommand{\cJ}{\mathcal{J}}
\newcommand{\cK}{\mathcal{K}}
\newcommand{\cL}{\mathcal{L}}

\newcommand{\cV}{\mathcal{V}}
\newcommand{\cW}{\mathcal{W}}

\newcommand{\bp}{{\bf p}}
\newcommand{\bq}{{\bf q}}

\newcommand{\by}{{\bf y}}

\newcommand{\bF}{{\bf F}}
\newcommand{\bG}{{\bf G}}

\newcommand{\bK}{{\bf K}}

\newcommand{\bP}{{\bf P}}
\newcommand{\bQ}{{\bf Q}}

\newcommand{\bW}{{\bf W}}
\newcommand{\bX}{{\bf X}}
\newcommand{\bY}{{\bf Y}}

\newcommand{\sub}{\subset}

\newcommand{\wed}{\wedge}
\newcommand{\com}{\!\circ\!}
\newcommand{\ten}{\!\otimes\!}

\newcommand{\gam}{\gamma}

\newcommand{\eps}{\epsilon}
\newcommand{\zet}{\zeta}
\newcommand{\tht}{\theta}

\newcommand{\lam}{\lambda}

\newcommand{\ome}{\omega}

\newcommand{\Lam}{\Lambda}
\newcommand{\Sig}{\Sigma}

\newcommand{\cw}{\vartheta}

\newcommand{\For}{{\Lambda}}
\newcommand{\Con}{{\mathcal{C}}}
\newcommand{\Hor}{{\mathcal{H}}}
\newcommand{\Var}{{\mathcal{V}}}
\newcommand{\Thd}{{\Theta}}

\title{{\bf A variational perspective on classical Higgs fields in gauge-natural theories \thanks{Research supported by the University of Torino, MiUR and partially (M.P.) by the University of Brno (CZ).}
}}
\author{{\normalsize 
M. Palese and E. Winterroth}
\\{\footnotesize Department of Mathematics,
University of Torino}
\\{\footnotesize Via C. Alberto 10, 10123 Torino, Italy}\\ 
{\footnotesize e--mails: 
{\sc 
[marcella.palese, ekkehart.winterroth]@unito.it}}}
\date{}
\begin{document}

\maketitle

\begin{abstract}
Higgs fields on gauge-natural prolongations of principal bundles are defined 
by invariant variational problems and related canonical conservation laws 
along the kernel of a gauge-natural Jacobi morphism.
\medskip


\noindent {\em keywords}: jet, gauge-natural bundle, conserved quantities, Cartan connection, Higgs field.
\end{abstract}

\section{Introduction}

In 1981 D.J. Eck framed natural and gauge (classical) Lagrangian field theories within a geometric functorial construction, called a gauge-natural bundle, whereby physical fields are assumed to be sections of bundles functorially associated with gauge-natural prolongations (also called Ehresmann prolongations \cite{Ehr55}) of principal bundles,
by means of left actions of Lie groups on manifolds, usually tensor spaces \cite{Ec81}. 
In fact, the study of jet prolongations of principal bundles with structure group a Lie group $\bG$ has revealed of fundamental importance in Physics. Such a prolongation however, as well known, is not a principal bundle, while the structure of a principal bundle is given to the so-called gauge-natural prolongation of a principal bundle \cite{Ehr55,KMS93}.
More precisely, we consider Lagrangian field theories which are assumed to be invariant with respect to the action of a gauge-natural group $W^{(r,k)}_{n}\bG$ defined as the semidirect product of a $k$-th order differential group of the base manifold with the group of $r$-th order $n$-th velocities in $\bG$, with $n=\dim \bX$ is the dimension of the basis manifold. Since the group $\Diff (\bX)$ {\em is not} canonically embedded into the group $\Aut(\bP)$ of {\em all} automorphisms of the underling principal bundle $\bP$, there is {\em a priori} no
natural way of relating infinitesimal gauge transfomations with 
infinitesimal base transformations, so that Lie derivatives of a gauge field with respect to infinitesimal base transformations could be defined neither in a natural nor, at least {\em a priori},  in a canonical way. 

The question of the existence of covariant canonically defined conserved currents is involved with such features of the Lie derivative \cite{PaWi04}. It is a well known fact that the covariance of the Lagrangian and thus of the Euler-Lagrange equations does not guarantee the corresponding covariance of Noether conserved quantities. Generally speaking, the fixing of a linear connection on the base manifold and of a principal connection on the principal bundle is needed in order to get covariant conserved quantities in gauge-natural field theories  (a 
global Poincar\'e--Cartan form can be defined only by fixing such a couple of connections; see \eg  \cite{FaFr03}). 
However, we found that a {\em canonical} determination of Noether conserved quantities, without fixing any connection {\em a priori}, is always possible on a reduced bundle of $W^{(r,k)}\bP$ determined by the original  $W^{(r,k)}_{n}\bG$-invariant variational problem. Connections can be characterized by means of such canonical reduction \cite{FFPW08,FFPW10,Winterroth07}.
This is due to the fact that bundles of fields associated with the class of principal bundles obtained as gauge-natural prolongations of principal bundles have a {\em richer} structure than the ones associated with principal bundles {\em tout court}.  

A variational problem on jets of fibered manifolds is invariant with respect to the finite order contact structure induced by jets: we shall consider finite order Lagrangian variational problems in terms of exterior differentials of forms modulo contact forms as framed in the context of finite order variational sequences \cite{Kru90}; moreover, in the category of variational sequences on gauge-natural bundles, the Lie derivative of sections of bundles is (up to a sign)  {\em the vertical part} w.r.t. the contact structure ({\em not} the
vertical component w.r.t. the projection on the basis manifold) of gauge-natural lifts of infinitesimal 
principal automorphisms \cite{PaWi03}.
In a previous paper \cite{PaWi07},
we characterized in the framework of finite order variational sequences the second
variation of a gauge-natural invariant Lagrangian of arbitrary order and used this characterization to specialize the Noether theorems and corresponding conserved currents, as well as  {\em generalized canonically defined Bergmann--Bianchi identities} for the existence of superpotentials \cite{Ber49}. 
We tackled the problem of canonical covariance of conserved quantities by using  variational
derivatives taken with respect to generalized vector fields which are chosen to be Lie derivatives of
sections of gauge-natural bundles, taken w.r.t.
gauge-natural lifts of infinitesimal principal automorphisms. 

The problem of investigating conservation laws associated with a group of tranformations obtained by substituting the parameters with arbitrary functions ({\em unendlichen kontinuierlichen Gruppe}) was tackled in 1918 by Emmy Noether \cite{Noe18}, who established, in this case, the existence of certain identity relations between Euler-Lagrange expressions and their derivatives as a consequence of the invariance of a Lagrangian with respect to such a wider group of transformations (Noether identities). 
In 1956 Ryoyu Utiyama  \cite{Uti56} considered some systems of fields invariant under a certain  group of transformations depending on $n$ parameters and, {\em postulating} the invariance of such systems under the wider group obtained substituting the parameters with a set of arbitrary functions, he introduced a new field with a definite type of interaction with the original fields defined by {\em a covariant derivative}. 

In this perspective, we shall consider the class of parametrized contact transformations defined by the gauge-natural functor: resorting to invariance properties, we define covariant derivatives of fields and new conservation laws, through the construction of a principal connection, satisfying a certain additional condition. In particular, we characterize canonical covariant Lagrangian conserved quantities in classical field theory in terms of Higgs fields on such gauge principal bundles having the richer structure of a gauge-natural prolongation.
Under this perspective, topological conditions for the existence of a Cartan connection on the principal bundle $W^{(r,k)}\bP$ turn out to be necessary conditions for the existence of global solutions of Jacobi equations associated with the existence of canonically defined global conserved quantities. As an outcome, the Lie derivative of fields is constrained and it is parametrized by a Higgs field $h$ defined by 
the space of Jacobi fields.

\section{Jacobi fields generating canonical conservation laws}

We recall some useful concepts of prolongations; for details see \eg \cite{KMS93,Sau89}.
Let $J_s\bY$ of 
$s$--jet prolongations of (local) sections
of a fibered manifold $\pi : \bY \to \bX$, with $\dim \bX = n$ and $\dim \bY = n+m$.
The natural fiberings $\pi^{s}_{s-1}$ are {\em affine} fiberings
inducing a  natural splitting
$J_{s}\bY \times_{J_{s-1}\bY}T^*J_{s-1}\bY =
J_s\bY \times_{J_{s-1}\bY}\left(T^*\bX\oplus V^*J_{s-1}\bY\right)$ (see \eg \cite{Kru90,Vit98}) which yields rising order decompositions: given a vector field $\Xi : J_{s+1}\bY \to TJ_{s+1}\bY$, $ T \pi^{s+1}_{s}\com \Xi \, \, = \Xi_{H} + \Xi_{V}$, where
$\Xi_{H}$ and $\Xi_{V}$ are  {\em the horizontal} and {\em the vertical part of $\Xi$}, respectively; for the exterior differential
on $\bY$, $(\pi^{r+1}_r)^*\circ d = d_H + d_V$, where $d_H$ and $d_V$ are
called the \emph{horizontal} and \emph{vertical differential}, respectively;  the {\em sheaf splitting}
$\Hor^{p}_{(s+1,s)}$ $=$ $\bigoplus_{t=0}^p$
$\Con^{p-t}_{(s+1,s)}$ $\wed\Hor^{t}_{s+1}$, where the sheaves $\Hor^{p}_{(s,q)}$ and
$\Hor^{p}_{s}$ of {\em horizontal forms} with respect to the 
projections $\pi^s_q$ and $\pi^s_0$, respectively, while $\Con^{p}_{(s,q)}
\sub \Hor^{p}_{(s,q)}$ and $\Con^{p}{_s} \sub
\Con^{p}_{(s+1,s)}$ are {\em contact forms}, \ie horizontal forms 
valued into $\cC^{*}_{s}[\bY]$. We put $\Hor^{p,}{_{s+1}^{h}}$ $\doteq$ $h(\For^{p}_s)$ for $0 < p\leq n$, where the projection on the summand of lesser contact degree
$h$ is {\em the horizontalization}.

By an buse of notation, denote by $d\ker h$ the sheaf
generated by the presheaf $d\ker h$ in the standard way.
We set $\Thd^{*}_{s}$ $\doteq$ $\ker h$ $+$
$d\ker h$. We have {\em the variational sequence}
$0 \to \R_{Y} \to \Var^{*}_{s}$,
where $\Var^{*}_s=\For^{*}_s / \Thd^{*}_{s}$, which is an exact resolution of the constant sheaf $ \R_{Y} $ \cite{Kru90}.
A section $E_{d\lam}\doteq \cE_{n}(\lam) \in \Var^{n+1}_{s}$ is the {\em generalized higher 
order Euler--Lagrange type morphism} associated with $\lam$.

Let $\bP\to\bX$ be a principal bundle with structure group $\bG$.
For $r\leq k$ integers consider the {\em gauge-natural prolongation of $\bP$} given by 
$\bW^{(r,k)}\bP$ $\doteq$ $J_{r}\bP \times_{\bX}L_{k}(\bX)$, 
where $L_{k}(\bX)$ is the bundle of $k$--frames 
in $\bX$ \cite{Ec81,KMS93}; $\bW^{(r,k)}\bP$ is a principal bundle over $\bX$ with structure group
$\bW^{(r,k)}_{n}\bG$ which is 
the {\em semidirect} product with respect to the action of $GL_{k}(n)$ 
on $\bG^{r}_{n}$ given by  
jet composition and $GL_{k}(n)$ is the group of $k$--frames 
in $\R^{n}$. Here we denote by $\bG^{r}_{n}$ the space of $(r,n)$-velocities on $\bG$.
Let $\bF$ be a manifold and $\zet: \bW^{(r,k)}_{n}\bG \times_{}\bF\to\bF$ be 
a left action of $\bW^{(r,k)}_{n}\bG$ on $\bF$. There is a naturally defined 
right action of $\bW^{(r,k)}_{n}\bG$ on $\bW^{(r,k)}\bP \times \bF$ so that
 we have in the standard way the associated {\em gauge-natural bundle} of order 
$(r,k)$: $\bY_{\zet} \doteq \bW^{(r,k)}\bP\times_{\zet}\bF$.
All our considerations shall refer to $\bY$ as a gauge-natural bundle as just defined.

Denote now by $\cA^{(r,k)}$ the sheaf of right invariant vector fields 
on $\bW^{(r,k)}\bP$. The {\em gauge-natural lift} is defined as the  functorial map 
$\mathfrak{G} : \bY_{\zet}  \times_{\bX} \cA^{(r,k)} \to T\bY_{\zet} \,:
(\by,\bar{\Xi}) \mto \hat{\Xi} (\by) $, where, for any $\by \in \bY_{\zet}$, one sets: $\hat{\Xi}(\by)=
\frac{d}{dt} [(\Phi_{\zet \,t})(\by)]_{t=0}$,
and $\Phi_{\zet \,t}$ denotes the (local) flow corresponding to the 
gauge-natural lift of $\Phi_{t}$. Such a functor defines a class of parametrized contact transformations.
This mapping fulfils the following properties (see \cite{KMS93}):
 $\mathfrak{G}$ is linear over $id_{\bY_{\zet}}$;
we have $T\pi_{\zet}\circ\mathfrak{G} = id_{T\bX}\circ 
\bar{\pi}^{(r,k)}$, 
where $\bar{\pi}^{(r,k)}$ is the natural projection
$\bY_{\zet} \times_{\bX} 
\cA^{(r,k)} \to T\bX$;
 for any pair $(\bar{\Lam},\bar{\Xi})$ $\in$
$\cA^{(r,k)}$, 
$\mathfrak{G}$ is a homomorphism of Lie algebras.

The  Lie derivative is a fundamental geometric object providing  information on how solutions of Euler-Lagrange equations  behave under the action of infinitesimal transformations (automorphisms) of the gauge-natural bundle. Let $\gam$ be a (local) section of $\bY_{\zet}$, $\bar{\Xi}$ 
$\in \cA^{(r,k)}$ and $\hat\Xi$ its gauge-natural lift. 
Following \cite{KMS93} we
define the {\em 
generalized Lie derivative} of $\gam$ along the vector field 
$\hat{\Xi}$ to be the (local) section $\pounds_{\bar{\Xi}} \gam : \bX \to V\bY_{\zet}$, 
given by
$\pounds_{\bar{\Xi}} \gam = T\gam \circ \xi - \hat{\Xi} \circ \gam$.
Due to the functorial nature of $\hat{\Xi}$, the Lie derivative of sections inherits some useful linearity properties and, in particular, it is an homomorphism of Lie algebras. In the view of Noether theorems, the interest of the Lie derivative of sections is due to the fact that, for any gauge-natural lift, we have 
$\hat{\Xi}_V = - \pounds_{\bar{\Xi}}$.
In the following we shall consider variational sequences on gauge-natural bundle $\bY$.

Let $\eta\in\Con^{1}_{s}\wed\Con^{1}_{(s,0)}\wed\Hor^{n,}{_{s+1}^{h}}$ and $\Xi$ a vertical vector field;
the morphism $E_{{j_{s}\Xi}\rfloor \eta} =  (\pi^{2s+1}_{s+1})^* {j_{s}\Xi}\rfloor \eta+F_{{j_{s}\Xi}\rfloor \eta}$ (with $F_{{j_{s}\Xi}\rfloor \eta}$ a local divergence) is a uniquely defined global section of $\Con^{1}_{(2s,0)}\wed\Hor^{n,}{_{2s+1}^{h}}$ (see \cite{Vit98}). 
Let $\lam$ be a Lagrangian, $\hat{\Xi}_{V}$ a generalized variation vector field and 
$\eta= hd\cL_{j_{2s}\bar{\Xi}_{V}}\lam$. Let us set $\chi(\lam,\hat{\Xi}_{V})$ $\doteq$  
$E_{j_{s}\hat{\Xi}\rfloor \eta}$. By resorting to functorial linearity properties of $\hat{\Xi}$ we define a 
{\em linear} morphism, the {\em gauge-natural generalized Jacobi 
morphism} associated with the Lagrangian $\lam$ and the variation vector field 
$\hat{\Xi}_{V}$, $\cJ(\lam,\hat{\Xi}_{V})$ $\doteq$
$E_{\cdot\rfloor \chi(\lam,\hat{\Xi}_{V})}$ \cite{PaWi03}.
It turns out that $\cJ(\lam,\hat{\Xi}_{V})$, the second variational derivative $\cL_{j_{s}\bar{\Xi}_{V}}\cL_{j_{s}\bar{\Xi}_{V}}\lam$ and the Hessian morphism $\mathfrak{H}(\lam, \hat{\Xi}_{V})$ $\doteq$ $ \hat{\Xi}_{V}\rfloor
\cE_{n}(\hat{\Xi}_{V}\rfloor\cE_{n}(\lam))$ are all representatives of the same equivalence class in a suitable variational sequence \cite{PaWi07}, thus characterizing $\cJ(\lam,\hat{\Xi}_{V})$ as a symmetric self-adjoint morphism.
The relevance of this property is concerned with important geometric aspects of the space 
$\mathfrak{K}\doteq \ker\cJ(\lam,\hat{\Xi}_{V})$. which defines generalized gauge-natural Jacobi equations, the solutions of which we call {\em generalized Jacobi vector fields} and characterize {\em canonical} covariant conserved quantities \cite{PaWi04}.

It is well known that the First Noether Theorem can be recasted by resorting to the {\em variational Lie derivative} of classes of forms represented the variational sequence:
$\cL_{j_{s}\Xi}\lam =
\cw (\lam,\pounds_{\bar{\Xi}} ) +
d_{H}\eps(\lam,\pounds_{\bar{\Xi}} ) $, where we put  $\cw(\lam,\pounds_{\bar{\Xi}} )   \doteq  -\pounds_{\bar{\Xi}} 
\rfloor \cE_{n} (\lam)$ and $\eps(\lam,\pounds_{\bar{\Xi}} )$ is a Noether current.
As usual,  $\lam$ is defined a
{\em gauge-natural invariant Lagrangian} if the gauge-natural lift
$(\hat{\Xi},\xi)$ of any vector
field $\bar{\Xi} \in \cA^{(r,k)}$ is a  symmetry for
$\lam$, \ie if $\cL_{j_{s}\bar{\Xi}}\,\lam = 0$.
It is remarkable that, in general, $\cL_{j_{s}\bar{\Xi}_V}\,\lam $ $\neq$ $ 0$.

As already mentioned the existence of a canonical global superpotential for $\eps(\lam,\pounds_{\bar{\Xi}})$ relies on  covariant Bergmann-Bianchi identities, which  can be proved to exist canonically only along $ \ker\cJ(\lam,\hat{\Xi}_{V})$ \cite{PaWi03}. Owing to the fact that they are Noether identities associated with the invariance properties of the Euler--Lagrange morphism $\cE_{n}(\cw)$, the kernel $\mathfrak{K}$, being the kernel of a Hamiltonian operator, can be characterized as a vector subbundle \cite{PaWi07,PaWi08a}. It is relevant for the theory of Lie 
derivative of gauge-natural fields that
 the intrinsic indeterminacy of conserved charges associated 
with gauge-natural symmetries of Lagrangian field theories is in this way solved. 

\section{Higgs fields on gauge-natural bundle}

By an abuse of notation, we denote by $\mathfrak{k}$ the Lie algebra of generalized Jacobi vector fields.  
Let $\mathfrak{h}$ be the Lie algebra of right-invariant vertical vector fields on $W^{(r+4s,k+4s)}\bP$. Now, let us assume that global solutions of generalized gauge-natural Jacobi equations exist; the Lie algebra $\mathfrak{k}$ is then characterized as Lie subalgebra of $\mathfrak{h}$; the Jacobi morphism self-adjoint  and $\mathfrak{k}$ is of constant rank; the  split structure $\mathfrak{h}=\mathfrak{k}\oplus \textstyle{Im}\,\cJ$ is well defined and it  is also reductive, being $[ \mathfrak{k},\textstyle{Im}\,\cJ]=\textstyle{Im}\,\cJ$ \cite{PaWi08b}. 
In particular, for each $\bp\in W^{(r,k)}\bP$ by denoting $\cW\equiv\mathfrak{h}_{\bp}$, $\cK\equiv\mathfrak{k}_{\bp}$ and $\cV\equiv \textstyle{Im}\,\cJ_{\bp}$ we have the reductive Lie algebra decomposition $\cW=\cK\oplus\cV$, with $[\cK,\cV]=\cV$. Notice that $\cW$ is the Lie algebra of the Lie group $W^{(r,k)}_{n}\bG$.

As a consequence of the fact that $\cK$ is a reductive Lie algebra of $\cW$, there exists an isomorphism between $\cV\equiv \textstyle{Im}\cJ_{\bp}$ and $T\bX$ so that $\cV$ turns out to be the image of an horizontal subspace. Thus we caracterize a principal bundle $\bQ\to\bX$, with $\textstyle{dim}\bQ=\textstyle{dim}\cW$, such that $\bX=\bQ/\bK$. 
The principal subbundle $\bQ\sub \bW^{(r,k)}\bP$ is such that $\cK = T_{\bq}\bQ/\bK$, where 
$\bK$ is the (reduced) Lie group of the Lie algebra $\cK$ is a reduced principal bundle.

In the following we shall omit the orders of a gauge-natural prolongation to simplyfy the notation.
The Lie group $\bK$ of the Lie algebra $\cK$ is in particular a closed subgroup of $W\bG$ ($\mathfrak{k}$ is a vector subbundle). We have the composite fiber bundle
\beq
W\bP\to W\bP/\bK\to\bX\,,
\eeq
where $W\bP\to W\bP/\bK$ is a pricipal bundle with structure group $\bK$
and 
$W\bP/\bK$ $\to$ $\bX$ is a fiber bundle associated with $W\bP$ with typical fiber $W\bG/\bK$, on which the structure group $W\bG$ acts on the left. 
Thus $W\bP/\bK =W\bP\times_{W\bG} W\bG/\bK \to\bX$ is a gauge-natural bundle functorially associated with $W\bP\times W\bG/\bK \to\bX$  by the right action of $W\bG$.
The left action of $W\bG$ on $W\bG/\bK$ is in accordance with  the reductive Lie algebra decomposition $\cW=\cK\oplus\cV$, with $[\cK,\cV]\sub \cV$. In fact the vector space $\cV=\textstyle{Im}\,\cJ_{\bp}=\cW/\cK$ 
carries the left action of $\cW$ given by the adjoint representation (since $[\cK,\cV]\sub\cV$ we have also 
$[\cW,\cV]=[\cK\oplus\cV,\cV] = [\cK,\cV] \oplus [\cV,\cV]  \sub\cV$).
 
We call a global section $h: \bX\to W\bP/\bK$ a {\em gauge-natural Higgs field}. Notice that it is a 
vector field lying in the image of the Jacobi morphism (recall that $\cW/\cK=\cV=\textstyle{Im}\,\cJ_{\bp}$), which in turn is an image of an horizontal subspace $T\bX$.  
The pull-back bundle $\bQ \doteq W\bP\ucar{W\bP/\bK} h(\bX)\sub W\bP/\bK = W\bP\ucar{W\bP/\bK} (h(\bX) \sub W\bP\times_{W\bG} W\bG/\bK)\to \bX$ is a reduced principal subbundle of $\bP$.
 Notice that a gauge-natural Higgs field is a global section of  $\hat{H}_p$, with $\bp\in \bQ$. 

Let $\ome$ be a principal connection on $\bW^{(r,k)}\bP$ and  $\bar{\ome}$ a principal connection  on the principal bundle $\bQ$ \ie a $\cK$-invariant horizontal distribution defining the vertical parallelism $\bar{\ome}: V\bQ\to \cK$ in the usual and standard way. It defines the splitting $T_{\bp}\bQ\simeq_{\bar{\ome}} \cK\oplus \hat{H}_{\bp}$, $\bp\in\bQ$. Since $\cK$ is a subalgebra of the Lie algebra $\cW$ and $dim\bQ=dim\cW$, it is defined a principal Cartan connection of type $\cW/\cK$, that is a $\cW$-valued absolute parallelism $\hat{\ome}: T\bQ\to \cW$ which is an homomorphism of Lie algebras, when restricted to $\cK$; preserving Lie brackets if one of the arguments is in $\cK$, and such that  it is an extension of the natural vertical parallelism, \ie 
$\hat{\ome} |_{VQ}=\bar{\ome}$. 
In \cite{FFPW10} we defined $\hat{\ome}$ as the restriction  to $T\bQ$ of the natural vertical parallelism defined by a principal connection $\ome$ on $W^{(r,k)}\bP$ by means of the fundamental vector field mapping $\ome: VW^{(r,k)}\bP\to \cW$. It satisfies
 $H_{\bp}\cap T_{\bp}\bQ=\emptyset$, with $\bp\in\bQ$  and it  is a connection on $\bW^{(r,k)}\bP=\bQ\times_{\bK}W^{(r,k)}_{n}\bG\to \bX$, thus a Cartan connection on $\bQ\to\bX$ with values in $\cW$. 
We notice that it splits into the $\cK$-component which  is a principal connection form on the $\cK$-manifold $\bQ$, and
the $\cV$-component which is a displacement form. In fact, being $\bK$ a reductive Lie subgroup of $\bW_{n}^{(r,k)}\bG$ the principal Cartan connection could be seen as a 
$\bK$-structure equipped  with a principal connection form $\eta= \textstyle{pr_{\cK}} \,\circ \hat{\ome}$ on $\bQ$ \cite{PaWi09}. 

A gauge-natural Higgs field, being  a global section of  $\hat{H}_p$, with $\bp\in \bQ$, is related with the displacement form defined by  the $\cV$-component of the Cartan connection $\hat{\ome}$ above.
The principal bundle $W\bP$ (with Lie algebra of the structure group $\cW$) admits the principal subbundle $\bQ$ (with Lie algebra of the structure group $\cK$); furthermore, the direct sum $\cW = \cK \oplus \cV$ is given, where $\cV$ is a subspace such that $\ad (g)(\cV)\sub \cV$, $g\in \bK$. Then the pull-back  by $h$ of the $\cK$ valued component of a $\cW$ valued pricipal connection $\ome$ on $W\bP$ onto the reduced subbundle $\bQ$ is the connection form of a principal connection on $\bQ$ \cite{MaSa00}.
 
Given the composite fiber bundle 
\beq
W\bP\to W\bP/\bK\to\bX\,,
\eeq
we have the exact sequence
\beq
0\to V_{W\bP/\bK}   W\bP \to VW\bP \to W\bP\times_{W\bP/\bK}  VW\bP/\bK  \to 0\,,
\eeq
where $V_{W\bP/\bK}   W\bP$ denote the vertical tangent bundle of $ W\bP \to W\bP/\bK$.
 Every connection on the latter bundle  determines a splitting 
 \beq
 VW\bP= V_{W\bP/\bK}   W\bP\oplus_{W\bP/\bK} \tilde{\ome}(W\bP\times_{W\bP/\bK}  VW\bP/\bK)\,.
 \eeq
 by means of which we can define the vertical covariant differential 
$ \cD: J^{1}W\bP\to T^{*}\bX\ten_{W\bP}  V_{W\bP/\bK}   W\bP$ which is related with the 
covariant differential on $W\bP_h$ relative to the pull-back  connection $h^*(\tilde{\ome})$.

\subsection{Higgs fields and the Lie derivative of classical fields}

Let us now consider briefly, as a work example, the case of Lie derivative of spinor fields (this example as been exploited under this new perspective in \cite{Winterroth07,PaWi08a}, then also in \cite{FFPW08} concerning canonical connections).

On a $4$-dimensional manifold admitting Lorentzian structures 
($\textstyle{SO}(1,3)^{e}$-reductions) 
 $\bX$ consider a $\textstyle{SPIN}(1,3)^{e}$-principal bundle 
$\pi: \Sig \to \bX$ and a bundle map inducing 
a spin-frame on $\Sig$ given by $\tilde{\Lam}: \Sig\to L(\bX)$ defining a
metric  $g$ {\em via} the reduced subbundle  $\textstyle{SO}(\bX,g)=\tilde{\Lam}(\Sig)$ of $L(\bX)$.
A left action $\rho$ of the group $W^{(0,1)} \textstyle{SPIN}(1,3)^{e}$ on 
the manifold $GL(4,\R)$ is given so that  the associated 
bundle $\Sig_{\rho}\doteq W^{(0,1)} \Sig\times_{\rho} GL(4,\R)$ is is a gauge-natural bundle of order $(0,1)$, the {\em bundle of spin-tetrads $\tht$} \cite{weinberg72}.
The induced metric is $g_{\mu\nu}=\tht^{a}_{\mu}\tht^{b}_{\nu}\eta_{ab}$, where $\tht^{a}_{\mu}$ are 
local components of a spin tetrad $\tht$ and $\eta_{ab}$ the Minkowski metric.
Let $\mathfrak{so}(1,3)\simeq \mathfrak{spin}(1,3)$ be the Lie algebra of 
$\textstyle{SO}(1,3)$. One can consider the left action of $W^{(1,1)} \textstyle{SPIN}(1,3)^{e}$ on
the vector space $(\R^{4})^{*}\ten \,\mathfrak{so}(1,3)$.
The associated bundle
$\Sig_{l}\doteq W^{(1,1)} \Sig\times_{l}((\R^{4})^{*}\ten \mathfrak{so}(1,3))
$ is a gauge-natural bundle of order $(1,1)$,
the {\em bundle of spin-connections $\ome$}.
If $\hat{\gam}$ is the linear representation of $\textstyle{SPIN}(1,3)^{e}$ 
on the vector space $\C^{4}$ induced by the choice of matrices $\gam$ we get a $(0,0)$-gauge-natural
bundle
$\Sig_{\hat{\gam}}\doteq \Sig\times_{\hat{\gam}}\C^{4}$, the {\em bundle of spinors}.
A spinor connection $\tilde{\ome}$ is defined in a standard way in terms of the spin connection.

In the following the Einstein--Cartan Lagrangian will be the base preserving morphism
$\lam_{EC}: \Sig_{\rho}\ucar{\bX}J_{1}\Sig_{l}\to \Lam^{4}T^{*}\bX$,
while the Dirac Lagrangian is the base preserving morphism 
$\lam_{D}: \Sig_{\rho}\ucar{\bX}\Sig_{l} \ucar{\bX}J_{1}\Sig_{\hat{\gam}}$ $\to$ $ \Lam^{4}T^{*}\bX$ (local expressions can be found \eg in \cite{FaFr03}). We assume that the total Lagrangian of a gravitational field
interacting with spinor matter is
$\lam=\lam_{EC}+\lam_{D}$.

Let now $\bar{\Xi}$ be a $\textstyle{SPIN}(1,3)^{e}$-invariant vector field on $\Sig$. 
The lagrangian $\lam$ is invariant with respect to any lift $\hat{\Xi}$ of $\bar{\Xi}$ to the total
space of the theory. 
By the First Noether Theorem a conserved  
Noether current  $\eps(\lam, \bar{\Xi})$ can be found such that the corresponding superpotential is 
$\nu(\lam,\bar{\Xi})\doteq -\frac{1}{2k}\bar{\Xi}_{v}^{ab}\eps_{ab}$, where $\bar{\Xi}^{a}_{v\,b}=\bar{\Xi}^{a}_{b} -\ome^{a}_{b\mu}\xi^{\mu}$ is the
vertical part of $\bar{\Xi}$ with respect to the spin-connection $\ome$. 

We mentioned that, without the fixing of a connection {\em a priori}, the existence
of {\em canonical} global  conserved quantities in
field theory is related with Noether identities:
 in \cite{Winterroth07} we found that this implies
$\bar{\Xi}_{v}^{ab}=-\tilde{\nabla}^{[a}\xi^{b]}$ (the so-called Kosmann lift \cite{Kos66}), where
$\tilde{\nabla}$ is the covariant derivative with respect to the standard transposed connection on
$\Sig_{\rho}$. The Kosmann lift can be characterized from a variational point of view: it is 
the only gauge-natural lift
which ensures the naturality condition 
$\cL_{j_{s+1}\hat{\Xi}_{H}}[\cL_{j_{s+1}\hat{\Xi}_{V}}\lam]\equiv 0$ (Noether identities) holds true. Along
such a lift not only the initial Lagrangian $\lam$ is by assumption invariant, 
but also its first variational derivative $\cw(\lam,\mathfrak{K})$ is it.  On the other hand, the Lie derivative of spinor fields can be written in terms of a canonical spinor-connection $\tilde{\ome}$
as follows:
\beq
\pounds_{\bar\Xi}\psi= 
\xi^{\mu}\der_{\mu}\psi +\frac{1}{4}\hat{\Xi}_{h\,ab}\gam^{a}\gam^{b}\psi - \frac{1}{4}
\nabla_{[a}\xi_{b]}\gam^{a}\gam^{b}\psi\,,
\eeq
where $\hat{\Xi}_{h}$ is the horizontal part of $\hat{\Xi}$ {\em with respect to the  spinor-connection} \cite{FFPW08}.

It is clear that we are here considering the reduction of the total principal bundle which is the underlying structure bundle of the theory.
Each global section $h$ of $W\bP/\bK\to \bX$ (recall: $\bK$ comes from gauge-natural Jacobi equations) enables one to define a vertical covariant differential, which is related with the vertical differential defined by the principal connection on the total gauge-natural prolongation $W\bP$, thus also with  spin and spinor connections induced functorially on the associated bundle, as just shown.
We get out as an outcome that the Lie derivative of fields is constrained and it is parametrized by a Higgs field $h$ defined by 
$\mathfrak{K}$.
In particular, the Kosmann lift to the total bundle of tetrads and spinors is associated with a {\em variational} Higgs field on a gauge-natural bundle.


\end{document}